\documentclass[]{spie}  

 \usepackage[T1]{fontenc}
\usepackage{lmodern}
\usepackage{amsmath,amsfonts,amssymb}
\usepackage{graphicx}
\usepackage{sidecap}
\usepackage{wrapfig}
\usepackage{subfig}
\usepackage[colorlinks=true, allcolors=blue]{hyperref}

\usepackage{afterpage}

\usepackage{siunitx}
\DeclareSIUnit{\cmps}{\cm\per\second}
\DeclareSIUnit{\mps}{\meter\per\second}
\DeclareSIUnit{\kmps}{\kilo\meter\per\second}
\DeclareSIUnit{\micrometer}{\micro\meter}
\DeclareSIUnit{\foot}{'}
\DeclareSIUnit{\inch}{"}
\usepackage{amsmath}

\title{MAROON-X: The first two years of EPRVs from Gemini North}

\author[a]{Andreas Seifahrt}
\author[a]{Jacob L. Bean}
\author[a]{David Kasper}
\author[b]{Julian St\"urmer}
\author[a]{Madison Brady}
\author[a]{Robert Liu}
\author[c]{Mathias Zechmeister}
\author[d]{Gu\dh mundur Stef\'ansson}
\author[e]{Ben Montet}
\author[f]{John White}
\author[f]{Eduardo Tapia}
\author[f]{Teo Mocnik}
\author[f]{Siyi Xu}
\author[g]{Christian Schwab}
\affil[a]{The University of Chicago, Chicago, United States}
\affil[b]{Landessternwarte Heidelberg, Heidelberg, Germany}
\affil[c]{Georg-August-Universit\"at G\"ottingen, G\"ottingen, Germany}
\affil[d]{Princeton University, Princeton, United States}
\affil[e]{University of New South Wales (UNSW), Sydney, Australia}
\affil[f]{Gemini Observatory, Hilo, United States}
\affil[g]{Macquarie University, Sydney, Australia}

\authorinfo{Further author information: E-mail: seifahrt@uchicago.edu}

\begin{document} 
\maketitle

\begin{abstract}
MAROON-X is a fiber-fed, optical EPRV spectrograph at the 8-m Gemini North Telescope on Mauna Kea, Hawai'i. MAROON-X was commissioned as a visiting instrument in December 2019 and is in regular use since May 2020. Originally designed for RV observations of M-dwarfs, the instrument is used for a broad range of exoplanet and stellar science cases and has transitioned to be the second-most requested instrument on Gemini North over a number of semesters. We report here on the first two years of operations and radial velocity observations. MAROON-X regularly achieves sub-m/s RV performance on sky with a short-term instrumental noise floor at the 30 cm/s level. We will discuss various technical aspects in achieving this level of precision and how to further improve long-term performance.
\end{abstract}


\section{INTRODUCTION}
\label{sec:intro}  
Extreme precision radial velocity (EPRV) observations remain an important tool for exoplanet science, particularly in conjunction with transit observations. From the combination of the radius and orbital period determined from the transit light curve and the mass and orbital eccentricity determined from radial velocities, the planet bulk density and insolation can be inferred. These are critical parameters both for putting statistical constraints on the mass-radius relationship and mass function of extrasolar planets and for vetting the best targets for atmospheric characterisation using ground- and space-based facilities, such as \textsc{JWST}.

Ultra-stabilized spectrometers such as HARPS and HARPS-N have paved the way to a whole new generation of EPRV instruments on both hemispheres, covering both the visual and infrared wavelength regime and operating on a wide range of telescope sizes. MAROON-X is part of this development, filling the gap of EPRV capabilities on a large-sized telescope open to the entire US community. 

While MAROON-X was specifically designed for following up small transiting planets around mid to late M dwarfs with sub-m\,s$^{-1}$ RV precision, the instrument has turned into a versatile workhorse for a broad range of exoplanet and stellar science cases. With a red-optical bandpass (500 -- 920\,nm), covered at a resolving power of R\,$\simeq$\,85,000, and simultaneous calibration and sky fiber, MAROON-X is now the second in-demand instrument on Gemini North.

MAROON-X was installed at the Gemini North Observatory in May 2019 and saw first light in September 2019. We reported about the commissioning and science verification results obtained in December 2019 in \citenum{mx2020}. Additional details about the instrument design can be found in \citenum{mx2016} and \citenum{mx2018}. 

In the following sections we provide a brief overview of the operational and RV performance found in the first two years of regular science observations, discuss the technical trade-offs that limit the performance as well as discuss ongoing upgrade projects. 

\section{MAROON-X Operational Performance}
Although MAROON-X is currently still classified as a Visiting Instrument\footnote{See \url{https://www.gemini.edu/instrumentation/current-instruments/maroon-x} for more information.}, it is  permanently installed at Gemini North.
MAROON-X participates in Gemini's queue observing scheme in three blocks of 1--5 weeks each during a typical semester. When MAROON-X is scheduled for observations, the fiber injection unit (FIU, aka frontend unit) is mounted at Gemini's bottom instrument port, a space shared with Gemini's facility instruments NIFS and NIRI. The fiber conduit is pulled up through the telescope pier where it is stored when MAROON-X is not scheduled for observations. The repeatability of the FIU boresight is within $1''$ on sky. Very little active re-alignment is required to reach the same level of repeatability for re-seating of the fiber connector.

Over time we found smaller than expected flexure between the telescope's peripheral wavefront sensors (PWFS) and the position of the science fiber even for large changes in elevation or cassegrain rotation angle. For example, during transit spectroscopy programs where we stay on target for up to 7\,hours, we found no need to re-acquire the target. Initial (telescope-side) target acquisition places the science target well within a radius of $<2''$ around the science fiber when target positions and proper motions are correctly entered in the observing tool (OT). The combination of pointing accuracy and low flexure eliminated the need for offloading offsets from the tip-tilt mirror in the FIU to the telescope guiding system. The software interface was subsequently re-written to eliminate any active control of the telescope from the MAROON-X instrument control software.

Focus stability is equally better than expected. Small adjustments are sometimes made at the beginning of a run but are only affecting the coupling efficiency of calibration light into the science fiber by a few percent. This is because science (downstream) and calibration (upstream) fiber have the same size and the image of the calibration fiber is re-imaged 1:1 onto the science fiber after passing twice through camera, ADC, and tip-tilt optics in the FIU, making this setup very sensitive to small amounts of de-focus. We have not found the need for regular re-focusing to improve the coupling efficiency of stellar light, partially due to the independent (and automatic) focus control of the telescope. During periods of exceptional seeing, we measure stellar FWHM down to $0.3''$ on the science target with the FIU guide camera, another indication of adequate focus.

The guide camera in the frontend is operated with a typical integration time of 1--10\,sec, depending on the brightness of the source. Only targets as faint as V=16mag require exposure times of up to 15\,sec for reliable guiding on the 1\% of light diverted from the target. The use of the three back-illuminated single-mode fibers to triangulate the position of the science fiber is working  well. The triangulated science fiber position as well as the actual target position and shape are logged at the guide rate in the telemetry stream. Thanks to the low flexure, FIU guiding losses can be tolerated for many minutes before systematic offsets are notable. 

Since we have never encountered cases of field stars moving over the sky fiber location, the instrument is now operated with the telescope cassegrain de-rotator disabled, i.e. with the frontend unit at a fixed parallactic angle. This also eliminates the need for fast ADC updates during meridian passage. After an initial operational phase of manual ADC settings for each target, the ADC is now automatically tracking the telescope elevation and requires no further input during the night. 

Due to the constraints of Gemini's queue scheduling operation, observations are taken with fixed exposure times, rather than a fixed SNR. Depending on effective seeing and cloud cover, variations in SNR between exposures of the same target do occur, but seem to have no significant impact on RV precision.

The fiber conduit entering the Pier Lab above the instrument's thermal enclosure is dragged up and down and rotated with the telescope motion\cite{mx2020}. Plexi-glass walls were installed in 2021 on two sides of the roof of the enclosure to prevent the fiber from slipping off the enclosure. This also gives the fiber conduit more room to uncoil, practically eliminating earlier incidents of loops in the fiber conduit being dragged up during telescope motions.

Technical time loss during the first two years of operation was near zero with no significant failures during observing runs. In contrast, a number of power failures and glycol cooling outages occurred when the instrument was not scheduled for observations. This led to uncontrolled warm-ups of the detector cryostats and interrupted the otherwise continuous operation of the instrument. In particular, the unexpected loss of observatory-side glycol cooling led to an overheating of the waterjackets of the Sunpower Stirling cryocoolers which subsequently caused an uncontrolled power on/off cycle which imparted a large amount of vibrations on the detector systems, in one case for many hours. While this and other events did not lead to substantial misalignment or physical damage to the instrument, it did interrupt the RV baseline of the instrument due to shifts in focus and thus color-dependent changes in the instrumental profile. The largest event caused a significant increase in the resolving power achieved with the red arm of the instrument, a rather fortuitous outcome, all things considered.

\begin{figure}[b!]
\centering
\includegraphics[width=0.99\textwidth,keepaspectratio]{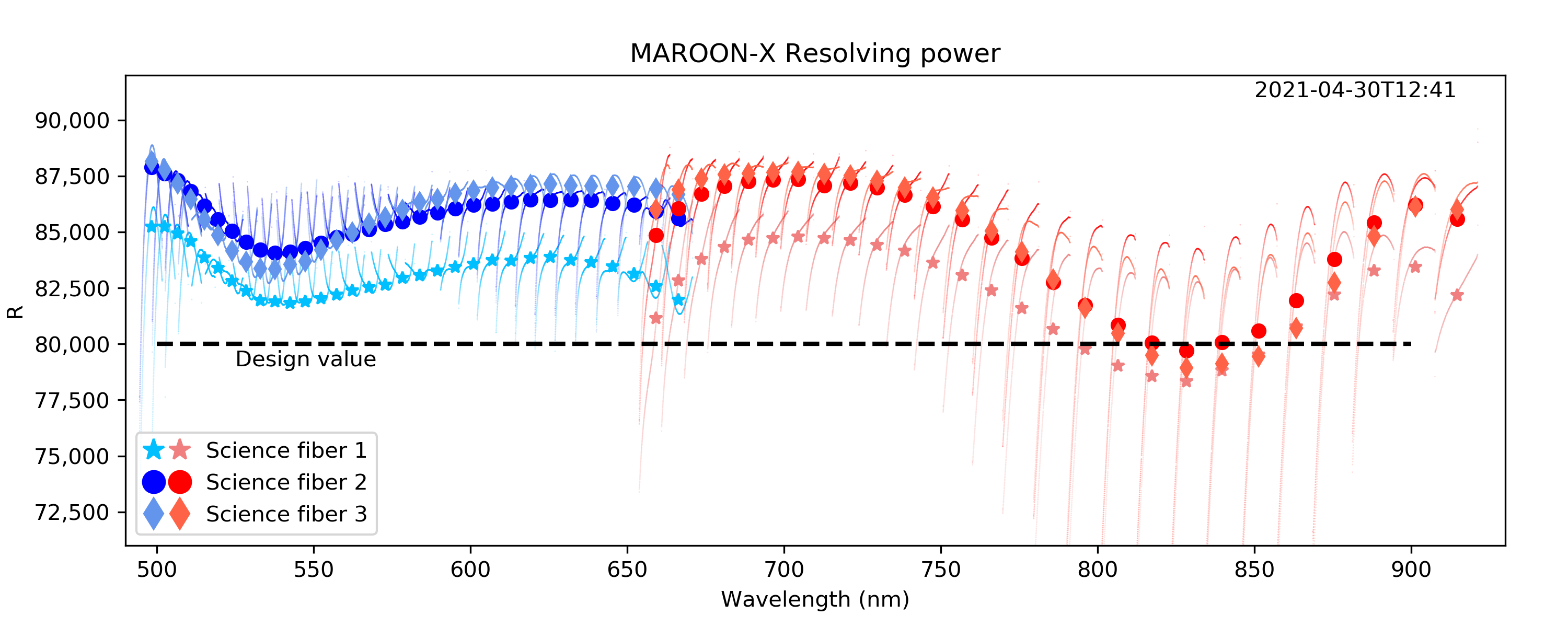}
\includegraphics[width=0.99\textwidth,keepaspectratio]{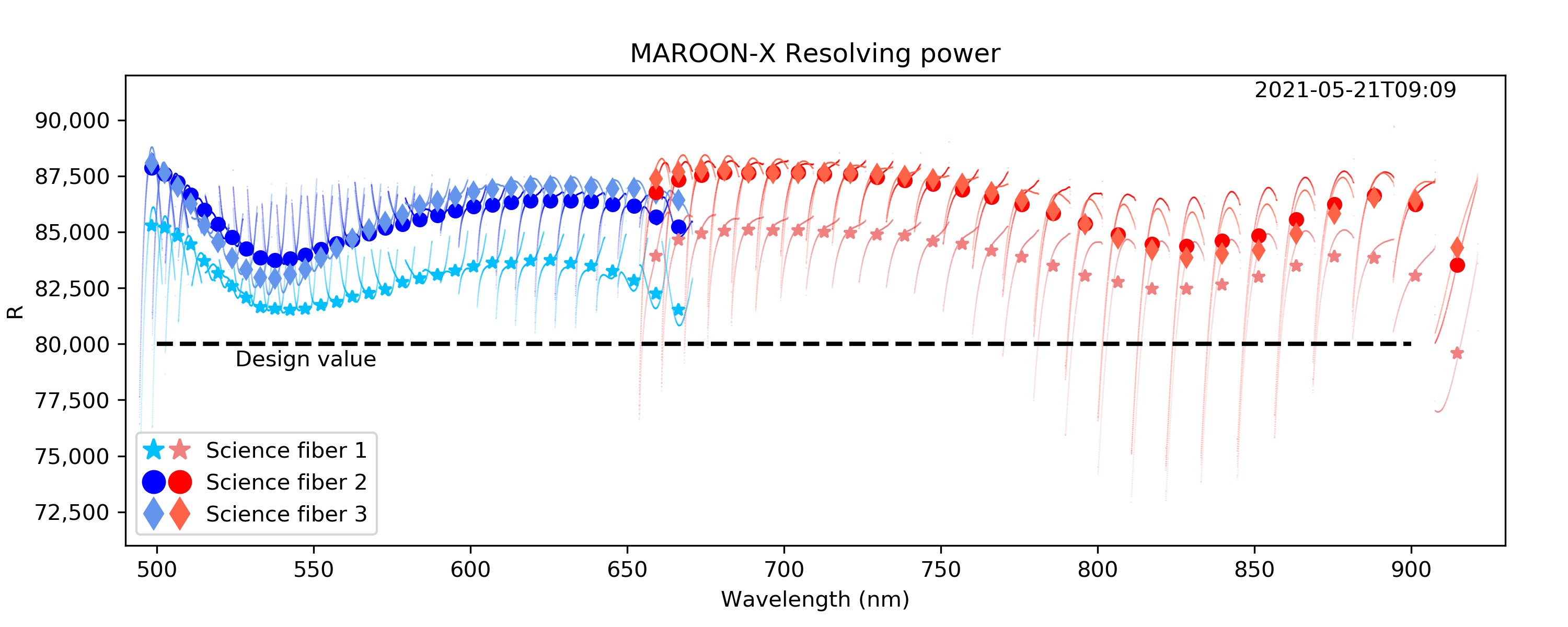}
\caption{\textbf{Change in the resolving power of MAROON-X due to a cooling failure on 2021-05-09 and vibrations imparted by the cryocoolers.} Shown are the values for three science fibers. Thin points show the raw values from each of the roughly 500 etalon lines per spectral order. Solid points mark the average value within the FSR of each order. Variations with wavelength and between fibers are driven by the complex aberration pattern of the MAROON-X camera optics.}
\label{fig:resolution_change}
\end{figure}

To prevent future cooling outages causing uncontrolled detector warm-ups, we installed a large coolant backup tank in the Pier Lab in the fall of 2021 which increases the buffer time for an observatory glycol outage from a mere 10\,min to over 12\,hrs, sufficient time for either a recovery of the observatory glycol supply system or a controlled warm-up of the detectors. Likewise, a number of software interlocks were installed to automatically shut off the cryocoolers in case of elevated vibration levels.

After two years of continuous 24/7 operation at 50\% power level, the NKT Compact supercontinuum source illuminating the etalon calibrator is not showing any signs of aging. Likewise, we are still operating with the same halogen flat and ThAr lamp since the commissioning of the instrument. Both lamps are only being used for about 2--3 $\times$ 15\,min each day during observing runs. 

The re-commissioning of the Rubidium tracing setup of the etalon calibrator was delayed after repeated altitude-related failures of the laser controller. The project was finally abandoned after it became clear that chromatic drifts ultimately limit the long-term use of the etalon and a Laser Frequency Comb could be ordered from Menlo Systems as an upgrade to the instrument (see Sec.~\ref{sec:longterm})

\section{MAROON-X RV Performance}
Despite being deployed at a large (8~m) telescope and offering a resolving power of R\,$\simeq$\,85,000 and a end-to-end efficiency peaking at 8\% under average seeing conditions, MAROON-X is an extremely compact instrument, with a beam diameter of only 100\,mm on a monolithic R4 echelle. This is possible thanks to an efficient $3\times$ pupil slicer \cite{slicer2016,mx2018} and the good seeing conditions on Mauna Kea, where the (70\%-ile) is reported at $0.75''$, allowing for efficienct observations with a small science fiber FOV of $0.77''$.  

\subsection{Instrument Design Impacts}
Owing to its compact size, MAROON-X could be build on a much smaller budget compared to typical instruments designed for 8-m class telescopes. Further acceleration of the development timescale was achieved by using a commercial vendor for the core echelle spectrograph. At the heart of MAROON-X is thus a modified KiwiSpec R4-100 spectrograph\cite{mx2016}. The main design feature of this spectrograph, aside from its strong asymmetric pupil compression of 33\%, is the placement of the optics. Only the main optics from the fiber slit plate up to the dichroic are in vacuum and referenced to the same optical bench, while the cross-disperser and camera arms as well as the detector systems are mounted outside of the vacuum chamber on a separate optical bench. 

This cost-effective design has the disadvantage of strong thermo-mechanical decoupling, aggrivated by the use of classical bench-mounted optics instead of a spaceframe design. Having the main dispersion direction in the vertical plane, small temperature differentials between the inside and outside of the vacuum chamber cause strong shifts in dispersion direction. Particularly the mounts of the detector systems, exposed to the air and surface temperatures outside of the vacuum chamber (inside the thermal enclosure) are a leading cause of drifts. While this design limitation was recognized from the start, the temperature sensitivity of the instrument turned out to be almost two orders of magnitude higher than anticipated, with sensitivities ranging from 1.8--7\,m\,s$^{-1}$\,mK$^{-1}$ \cite{mx2018}. This unfavorable feature has to be compensated by an optimized calibration strategy and by reducing the temperature fluctuations to a minimum.

\subsection{Instrument Calibration}\label{calibration}
It has been recognized for quite some time now that the classical calibration source for echelle spectrographs - the ThAr HCL - does not provide enough information content and spectral fidelity to be a viable calibration source for sub-\,m\,s$^{-1}$ RV measurements on its own. ThAr HCLs suffer from uneven spectral line density, high line contrast ratios (both between Thorium lines and the Thorium and fill gas), line blends, as well as from aging effects. For a discussion about the impact of ThAr based wavelength and drift solutions on ultra-stable spectrographs, see, e.g., \citenum{Coffinet}, \citenum{Cersullo}, \citenum{Schmidt2021}, and \citenum{Dumusque_HARPSN}. Compounding these problems are the sourcing limitations for pure Thorium and resulting contamination of the spectrum with ThO$_2$ lines\cite{dirt} in ThAr HCLs commercially available today. 

To overcome these limitations, we use a temperature- and pressure-stabilized Fabry-P\'erot etalon as the main calibrator for MAROON-X. Illuminated through a single-mode fiber by a NKT Compact supercontinuum laser, the etalon produces a comb-like spectrum of spectrally unresolved emission lines over the full wavelength range of MAROON-X\cite{stuermer2016,stuermer2017}.

Emission lines, regardless whether they are from the etalon or the ThAr HCL are 1D extracted and fitted with a line profile model that combines a box (representing the rectangular fiber) convolved on either side with a Gaussian profile (representing the optical aberrations of the spectrograph). This model fits the spectrally unresolved etalon lines extremely well and allows a direct measurement of changes in the line profile. We use the same model, with line profile parameters constrained by the etalon, to fit the ThAr lines. 

\begin{figure}[b!]
\centering
\includegraphics[width=0.99\textwidth,keepaspectratio]{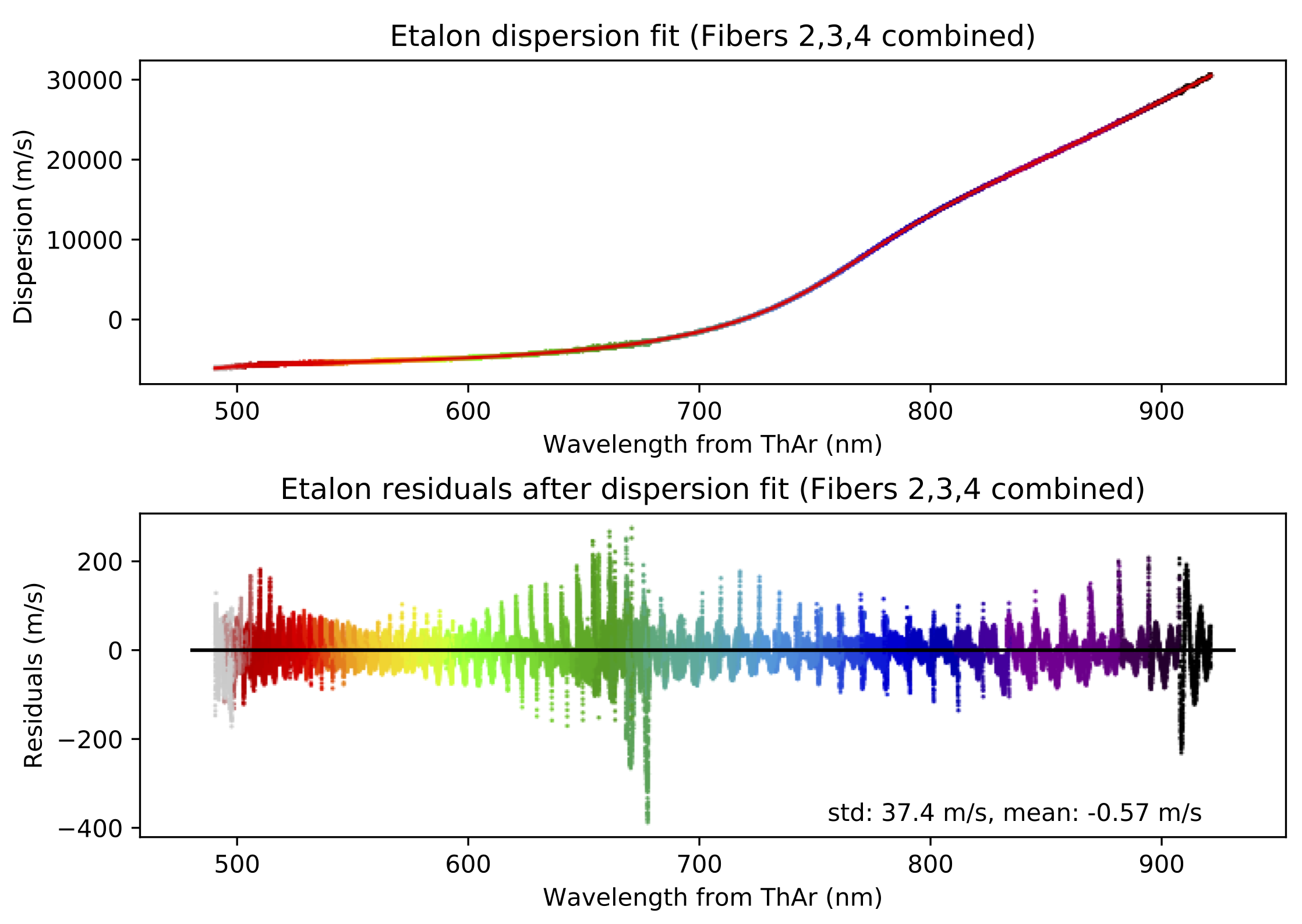}
\caption{\textbf{Etalon dispersion (ideal vs. real wavelengths of the etalon lines) based on thin film calculations of the coating stack model of the etalon mirrors.} \textit{Top:} Spline-based model after fitting to the etalon line positions on a ThAr-based wavelength solution using 2D Legendre polynomials of 8th and 6th degree in x and order number (y), respectively. The difference between initial dispersion model and observation never exceeds 10\% of the FSR of the etalon before the fit, guaranteeing a unique identification of etalon interference order numbers without discontinuities. \textit{Bottom:} Residuals to the fit highlighting the systematic errors in the ThAr-based wavelength solution. 100\,m\,s$^{-1}$ equals roughly 1\% of the FSR of the etalon at the middle of the instrument bandpass. Different colors indicate the different spectral orders of MAROON-X.}
\label{fig:dispersion}
\end{figure}

After an initial polynomial wavelength solution based on the ThAr spectrum, we can unambiguously identify the mode number of each etalon line as well as the as-built thickness of the etalon spacer based on a first guess for the etalon dispersion from thin film calculations of the etalon mirror coating stack. With these values we then re-fit the etalon dispersion using a cubic spline with knot locations strategically positioned to resolve the expected curvature of the dispersion to sub-\,m\,s$^{-1}$ precision, while at the same time preventing residuals of the ThAr-based 2D polynomial wavelength solution to be imprinted in the dispersion solution (see Fig.~\ref{fig:dispersion}). 

Due to the high pupil compression factor of the KiwiSpec R4-100, the optical aberration pattern is highly variable across the echellogram (see, e.g., Fig.~5 in \citenum{mx2016}). The high frequency of the spatial variability poses a problem to classical polynomial based wavelength solutions when enough information content is present to resolve the minute distortions in the pixel mapping of the wavelength scale. For example, a 2D Legendre polynomial was 'maxed out' at 8th and 6th degree in x and y, respectively, when using only ThAr lines for the wavelength solution. That is, a further increase in the polynomial degrees did not lead to a reduction in the residuals, even with careful outlier rejection. The residuals appear random with a typical rms of 160\,m\,s$^{-1}$. 

\begin{figure}[b!]
\centering
\includegraphics[width=0.9\textwidth,keepaspectratio]{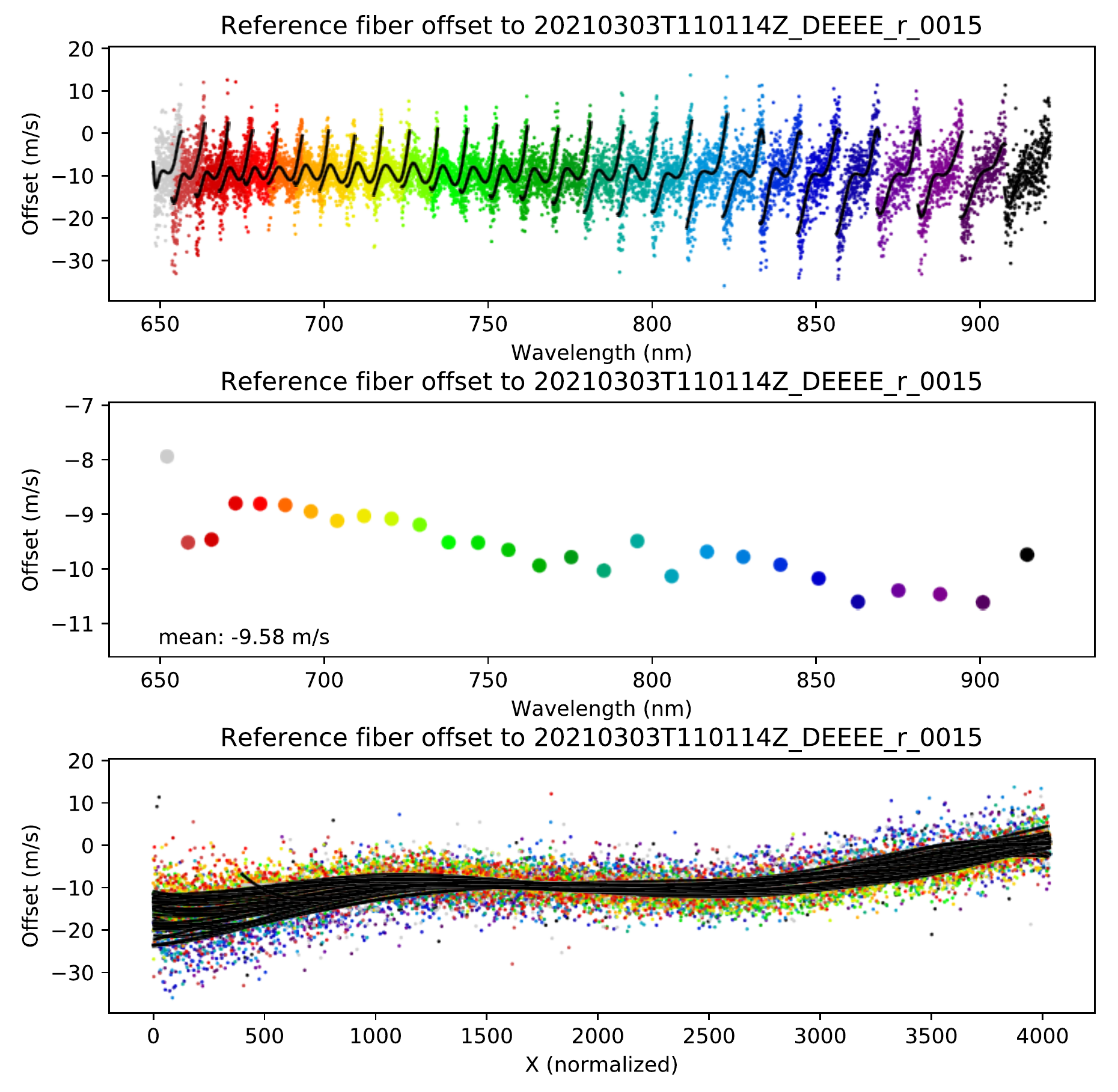}
\caption{\textbf{Example of an instrument drift correction derived from the etalon line position changes in the simultaneous calibration fiber} between a master etalon frame and a science frame. The time difference between the frames is only 24\,min, showing an extreme example of instrument drift and its correction. The mean instrument drift in the red arm is -9.6\,m\,s$^{-1}$, but there is significant structure, both in dispersion direction and in cross-dispersion direction that is not captured by a single drift value. \textit{Top:} Raw data (colored by order) and spline-based fit for the red arm. \textit{Center:} Average drift value in each order. \textit{Bottom:} Same as in the top plot but plotted over detector pixel position along each order.}
\label{fig:simcal}
\end{figure}

When using the etalon lines for the wavelength solution, the degrees of the polynomial can be increased as the local information density and fidelity is drastically larger than for the ThAr data. For example, on the initial (ThAr-based) polynomial wavelength solution, the etalon lines only showed residuals with a rms of under 40\,m\,s$^{-1}$, a reduction in comparison to the ThAr lines of a factor of four without any change to the underlying wavelength solution. Higher order polynomials continuously lower the residuals, but systematics in the form of ripples in the residuals remain both as a function of detector pixel (in dispersion direction) and in wavelength (in cross-dispersion direction). Systematics were still present when the polynomial reached its numerical stability limit around 20th degree in both directions. 

\begin{figure}[b!]
\centering
\includegraphics[width=0.9\textwidth,keepaspectratio]{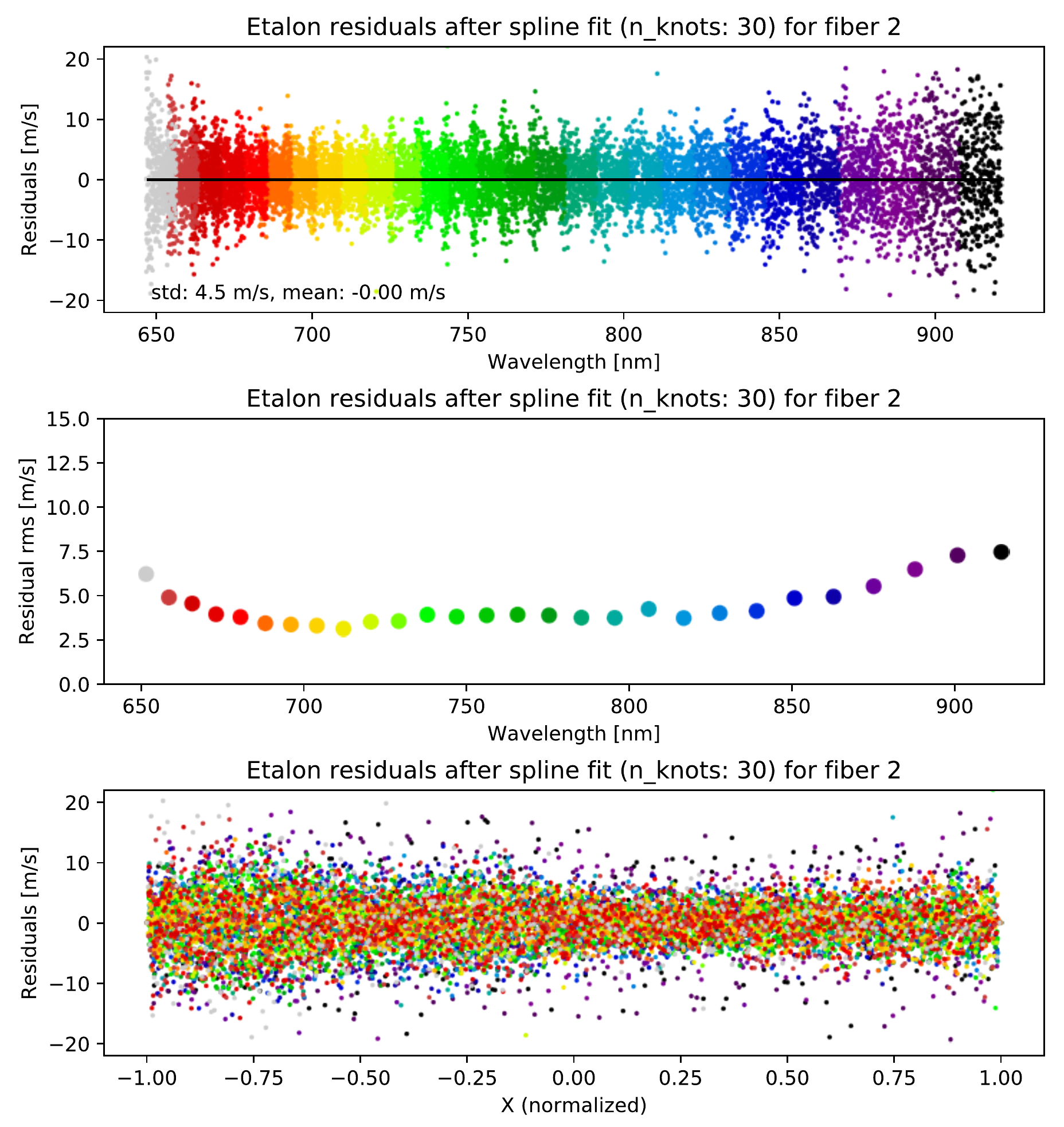}
\caption{\textbf{Example of the residuals of an etalon-based wavelength solution after applying the instrument drift correction shown in Fig.~\ref{fig:simcal}}. Each order is fitted by a cubic spline with 30 knots. The distribution of the residuals is entirely consistent with photon noise. \textit{Top:} Raw data (colored by order). \textit{Center:} Average rms in each order. \textit{Bottom:} Same as in the top plot but plotted over detector pixel position along each order.}
\label{fig:solution}
\end{figure}

We thus decided to use a cubic-spline based solution separately for each echelle order with 30 evenly spaced knots per order, just enough to remove the systematics present in polynomial-based solutions but preventing overfitting. The typical rms of the etalon lines on the new spline-based wavelength solution varies from 3--10\,m\,s$^{-1}$ and is photon-noise limited (see Fig.~\ref{fig:solution}).

Based on this initial solution we track the instrument drift over time by keeping the physical parameters of the etalon (spacer thickness and dispersion parameters) constant.

Each science frame taken with MAROON-X has an etalon spectrum in a dedicated fiber for simultaneous calibration (aka simcal fiber). A continuous ND filter wheel is set at the beginning of each exposure to allow for a constant flux level of the etalon spectrum independent of the exposure time. Daytime calibrations include dark frames with the same exposure times and etalon illumination level to subtract the faint extended wings of the etalon lines spilling over into the science fibers. 

Due to the imbalance in the etalon illumination, the red arm receives more etalon flux than the blue arm. Very bright science targets with significant blue flux (e.g., stars of solar or earlier type) can thus provide enough straylight to skew the fit of the etalon lines at the bluest orders in the blue arm at the few m\,s$^{-1}$ level. Global 2D straylight fitting and subtraction using the inter-order and inter-fiber space allows for a correction of this effect.

At the beginning and end of each night, a 'master etalon' spectrum is taken with etalon lines in both the science and simcal fibers. Comparing the pixel positions of the etalon lines in the simcal fiber between the science frame and the nearest master etalon frame provides a measure for the instrument drift between the two frames. This drift is than applied to the etalon positions in the science fibers of the master etalon frame which are then subsequently used to calculate a new wavelength solution for each science frame, effectively combining wavelength and drift solutions for each exposure. 

We found the need to strike a balance between spectral granularity of the drift solution and information content limited by photon noise and outliers from cosmic ray hits. Applying the offsets on individual lines appeared to introduce too much noise. Computing a order-based mean of the drift did not capture the apparent non-linear changes in dispersion. A practical solution was found by fitting the difference in the simcal etalon positions between the master etalon exposure and the science exposure with a cubic spline with three knots along each order, effectively spatially filtering the information, see Fig.~\ref{fig:simcal}. Applying the resultant spline to the wavelength solution derived from the master etalon frame (see Fig.~\ref{fig:solution}) leads to a robust drift solution that changes globally only at the 10\,cm\,s$^{-1}$ level when choosing different master etalon frames for the same science frame.

All science observations conducted since the start of regular operations in May 2020 have been reduced using this recipe for the drift solution of the spectrograph.

\subsection{Short-term on-sky RV Precision}\label{sec:short-term}

Radial velocity analysis of the stellar spectra uses the SERVAL package\cite{SERVAL} in a python3 implementation tailored to the data format of MAROON-X. The three science fibers can either be analyzed separately, or after combining them into a master science spectrum, a step that involves sub-pixel re-binning and spline interpolation based on the different wavelength solutions for each fiber. The latter is our standard approach and has been vetted for a number of targets to produce the same RVs as for the individual fibers.   

Observations of quiet RV standard stars as well as the first science results in early 2020 were encouraging, delivering on-sky RV residuals at the 30\,cm\,s$^{-1}$ level over timescales of two weeks\cite{mx2020}. We find that we can reproduce this level of on-sky precision over the short time scales of each observing run for the last two years and are intrinsically limited in assessing the on-sky instrumental stability floor by the activity levels of the stars.

\begin{figure}[t!]
\centering
\includegraphics[width=0.99\textwidth,keepaspectratio]{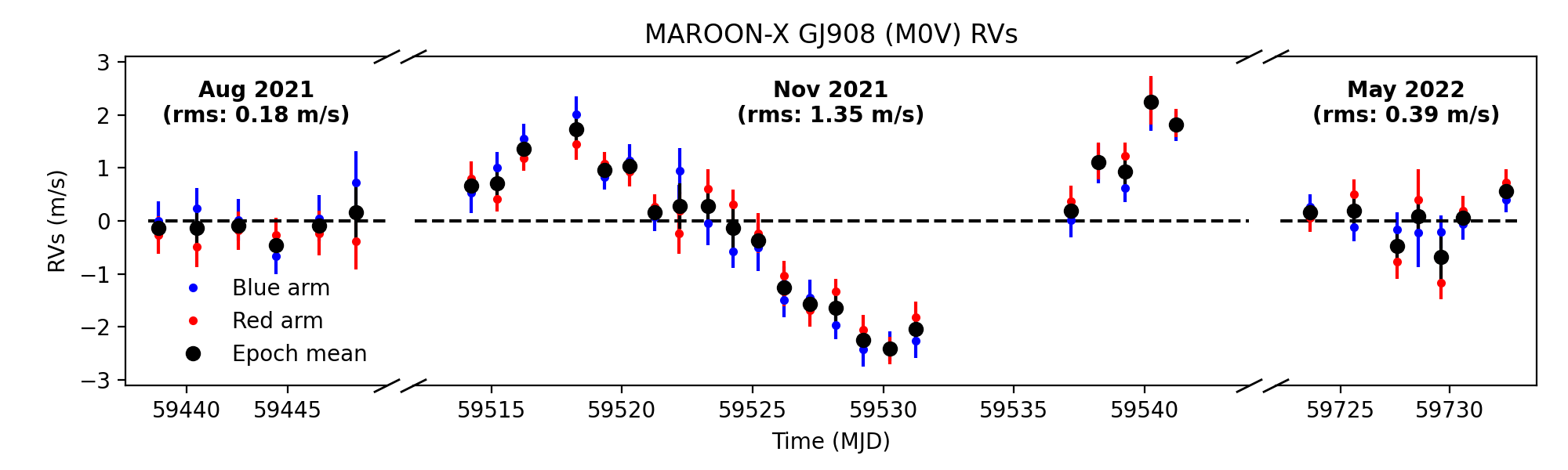}
\caption{\textbf{RVs of GJ908 (M0V), one of our RV standard stars observed in three runs in 2021 and early 2022 seen in and out of a period of activity.} GJ908 has shown RV rms values of well under 30\,cm\,s$^{-1}$ since MAROON-X science observations began in May 2020. During the last observing run in 2021 it suddenly showed signs of activity which seem to have mostly disappeared in May 2022.}
\label{fig:GJ908}
\end{figure}

GJ486b, our first science result contributed to \citenum{Trifonov}, showed a rms to the orbit fit of 44\,cm\,s$^{-1}$ for RVs from MAROON-X's red arm, without any activity modeling or de-trending. When combining the blue and red arms and binning the data over 30 minutes the RMS falls to $<$30\,cm\,s$^{-1}$. Observations a year later showed a similar level of performance\cite{Caballero}.

GJ908, a M0V star we found to be exceptionally quiet in early 2020 with RV rms of $<$30\,cm\,s$^{-1}$, remained at this level of stability until November 2021, when it suddenly showed a sinusoidal RV curve with a period of 23.7d, and a RV semi-amplitude of K = 1.75 m\,s$^{-1}$, which when subtracted leaves residuals of 34\,cm\,s$^{-1}$ for the red arm and 45\,cm\,s$^{-1}$ for the blue arm. During our latest observations in May 2022, the activity has come down almost to previous levels (see Fig.~\ref{fig:GJ908}). 

GJ908 illustrates the importance of observing multiple standard stars and to coordinate between different EPRV instruments to disentangle instrumental effects from stellar activity and from previously undiscovered low-mass planets around those stars. 

\begin{SCfigure}[][b!]
\centering
\includegraphics[width=0.60\textwidth,keepaspectratio,clip]{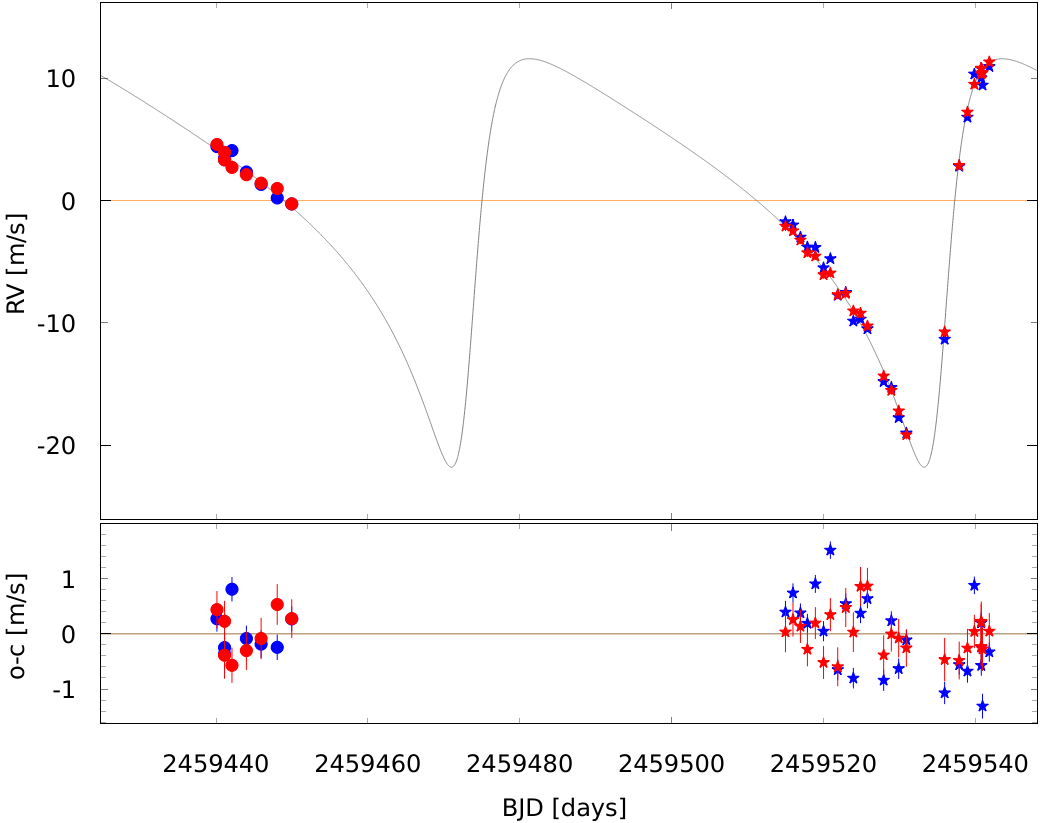}
\caption{\textbf{RVs of HD3651, a chromospherically quiet K1V star with a Saturn-mass planet in a highly eccentric orbit\cite{Brewer}}. We find a rms to the orbit fit of 38\,cm\,s$^{-1}$ for the red arm and 63\,cm\,s$^{-1}$ for the blue arm during two observing campaigns of two and four weeks in August and November 2021, respectively.}
\label{fig:HD3651}
\end{SCfigure}

An interesting approach to exclude the latter is HD3651, a chromospherically quiet K1V star with a highly eccentric Saturn-mass planet. The planet dynamically inhibits other stable planetary orbits. This system was first suggested by \citenum{Brewer} as an ideal RV benchmark. HD3651 is currently monitored by EXPRES\cite{EXPRES} and NEID\cite{NEID} and we started observing it in August 2021. The first results show residuals with an rms of 38\,cm\,s$^{-1}$ for the red arm and 63\,cm\,s$^{-1}$ for the blue arm of MAROON-X for two observing runs in August and November 2021 (see Fig.~\ref{fig:HD3651}).

\begin{figure}[t!]
\centering
\includegraphics[width=0.75\textwidth,keepaspectratio]{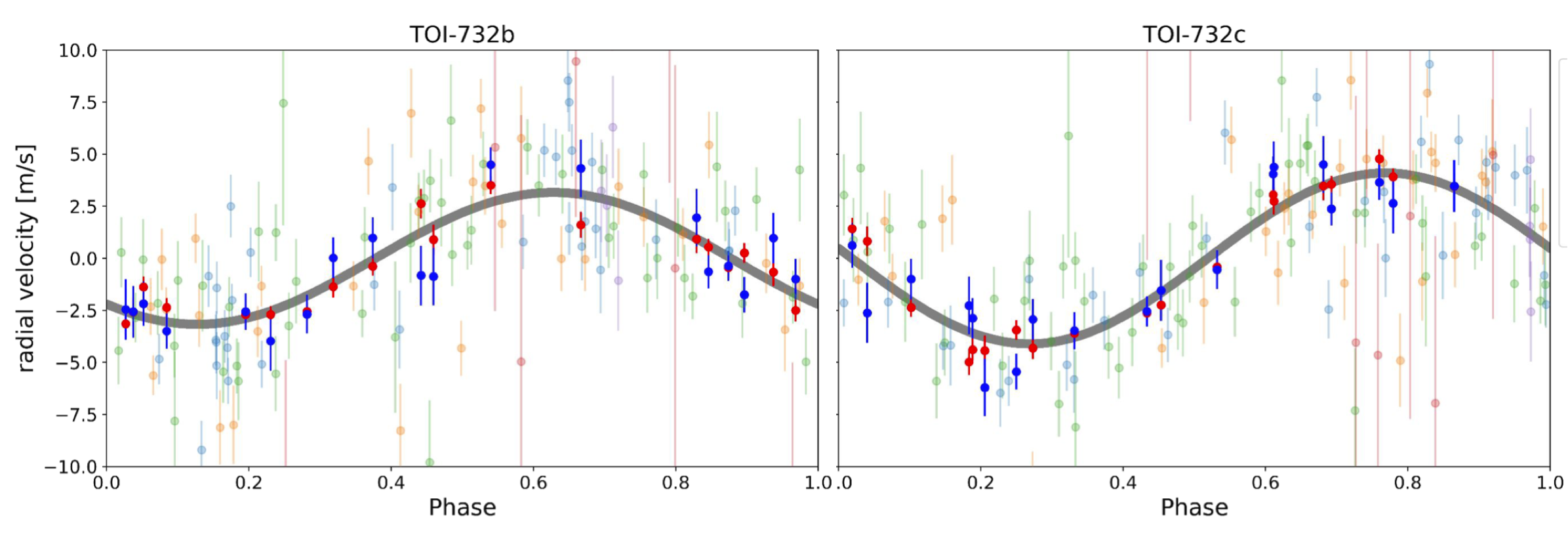}
\includegraphics[width=0.24\textwidth,keepaspectratio,trim={0cm -1cm 0cm 0cm}]{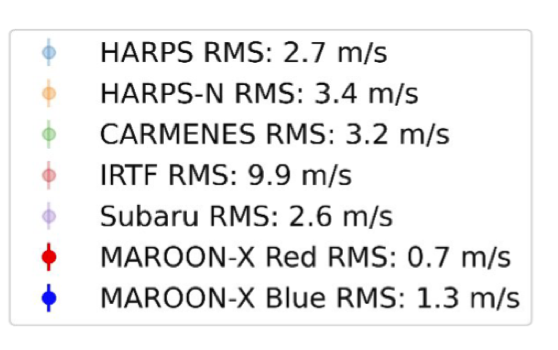}
\caption{\textbf{RVs of TOI-732 taken in early 2021 revealing the orbits of two transiting planets}, TOI-732b (left), a super-Earth and TOI-732c (right), a sub-Neptune in a 0.8\,d and 12.2\,d orbit, respectively. TOI-732 is a M4V star with V=14.7\,mag. MAROON-X RV measurements have much lower RV residuals than other datasets and a mass measurement of both planets to better than 10\% was achieved from only 19 epochs.}
\label{fig:TOI732}
\end{figure}

The demonstrated excellent short-term instrumental RV stability of MAROON-X and the fact that many M dwarfs have activity levels well below 1\,m\,s$^{-1}$ on short timescales opens up the possibility of detecting very small planets on orbits out to the distance of the circumstellar habitable zone with intensive observational campaigns. 

In Fig.~\ref{fig:TOI732} we show an example of an ongoing observing campaign to obtain homogeneous RV measurements to determine the masses of planets around M dwarfs discovered by TESS within 30\,pc of the sun. The
program aims at mass errors of 10\%, with a goal of 5\%, in order to place statistical constraints on the mass-radius relationship and mass function for M dwarf planets, to detect additional planets, and to provide target validation for atmospheric characterization with JWST. 

\subsection{Long-term RV Performance}\label{sec:longterm}

The long-term (months to years) RV performance of MAROON-X is harder to quantify. For the short-term performance demonstrated in Sec.~\ref{sec:short-term} we rely on the stability of the etalon calibrator to measure and compensate for the instrumental drift. Combining RVs from multiple runs at the same level of short-term precision is challenging. In the absence of the independent mono-chromatic zero-point measurements of the etalon provided by the Rb tracking setup\cite{stuermer2016,stuermer2017}, we have to rely on ThAr HCL data for absolute reference. 

After careful vetting of a NIST-based linelist, we find we can measure the instrumental RV drift of MAROON-X \textit{globally} with a precision of 20\,cm\,s$^{-1}$ in each arm. That is, when combining all orders in each arm, the combined information content allows to measure a global instrumental drift to this precision. These results depend on the removal of chromatic RV drifts with the etalon (i.e. the standard instrument drift correction as outlined in Sec.~\ref{calibration} above), which means that ThAr spectra must be taken like any other science spectrum, with etalon light in the simcal fiber. Bracketing etalon exposures around the ThAr frame show a marked increase in the uncertainty of the ThAr drift measurement by a factor of 2--3. In between runs, when MAROON-X is not scheduled for observations, the FIU is removed from the telescope and only the simcal fiber can be illuminated, excluding simultaneous measurements of ThAr and etalon calibrators and limiting precision ThAr measurements to times of active observing runs. 

When analyzing the long-term behavior of the ThAr data on the etalon-derived drift solution, we find an average drift of the etalon over time of 2.6\,cm\,s$^{-1}$\,d$^{-1}$, significantly lower than the 12\,cm\,s$^{-1}$\,d$^{-1}$ found in the lab for the same etalon\cite{stuermer2016}. This could be an effect of a slowed down shrinkage rate of the zerodur spacer over the 6 years between the measurements, or due to parasitic effects from the etalon mirror coatings.

Due to the numerous interruptions of the instrument baseline from power and cooling outages over the first two years, instrumental line profile changes occurred which are chromatic in nature and difficult to separate from chromatic drifts of the etalon itself. For example, the most extreme event leading to a notable focus shift of the instrument on 2021-05-09 (see Fig.~\ref{fig:resolution_change}), is clearly visible in the etalon data at various levels, from a dramatic change in the instrument's absolute RV zeropoint of over 1 km\,s$^{-1}$, to changes in the instrumental line profile (IP) - both its width (resolving power) and its shape. While even large changes in the RV zero point can be corrected to better than 1\,m\,s$^{-1}$, the changes to the IP have different impact on the calibration source and on stellar data and lead to unrecoverable RV offsets. 

\begin{figure}[t!]
\centering
\includegraphics[width=0.32\textwidth,keepaspectratio,trim={12.8cm 0cm 0cm 0cm},clip]{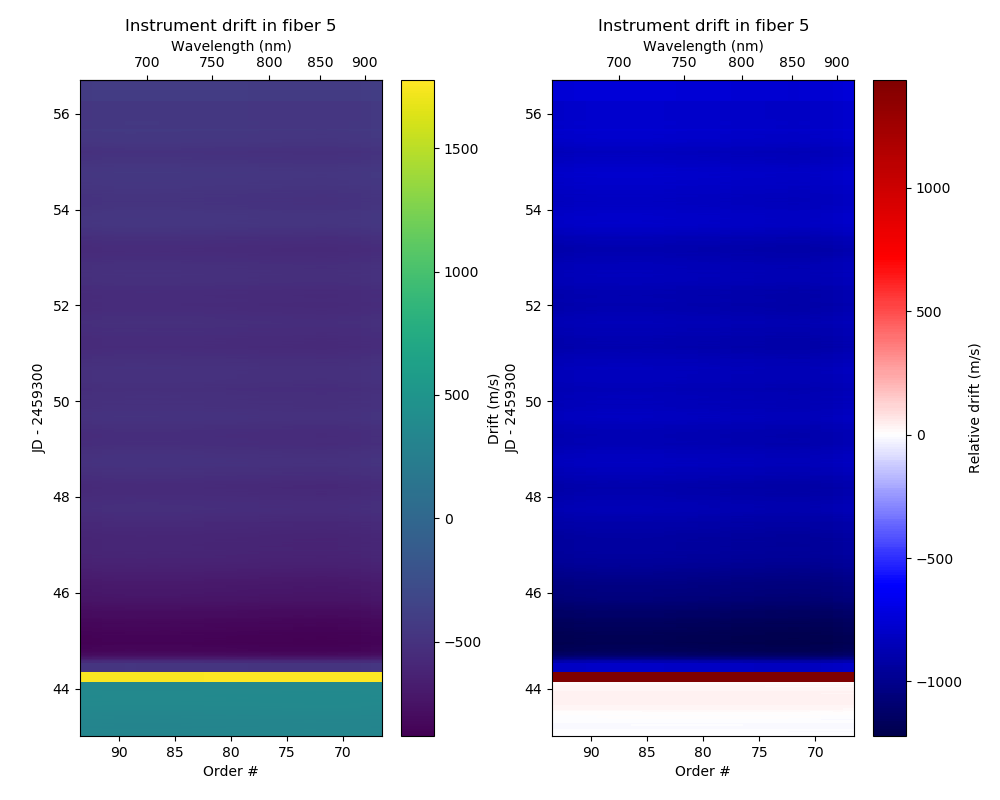}
\includegraphics[width=0.32\textwidth,keepaspectratio,trim={12.8cm 0cm 0cm 0cm},clip]{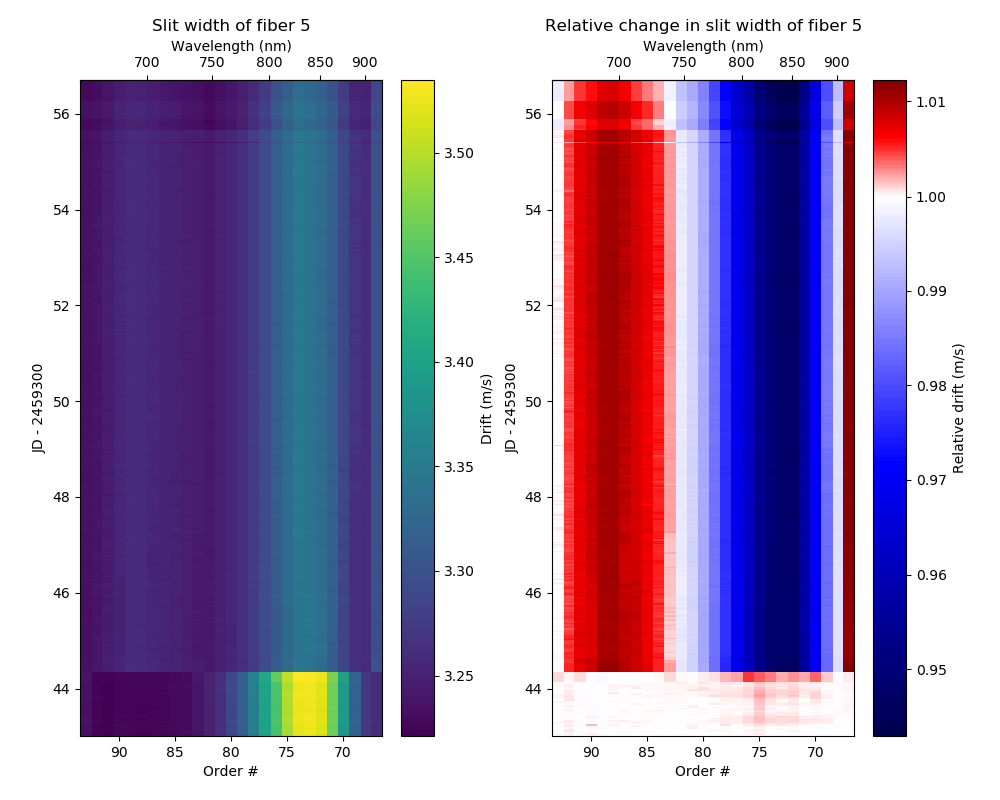}
\includegraphics[width=0.32\textwidth,keepaspectratio,trim={12.8cm 0cm 0cm 0cm},clip]{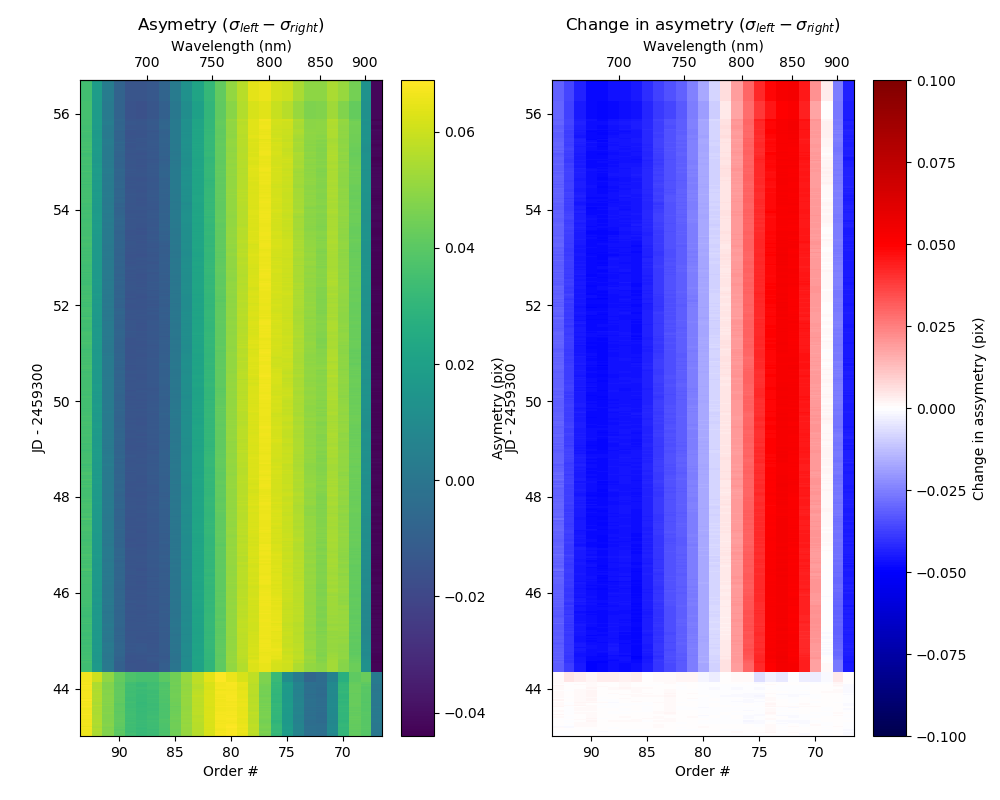}
\caption{\textbf{Relative change in instrument RV zeropoint (left), slit width (center), and symmetry of the line profile (right) for the event on 2021-05-09 as measured with the etalon}. Time is on the vertical axis, with the event occurring on JD 24593\textbf{44}. Compare to the resolving power changes in the red arm shown in  Fig.~\ref{fig:resolution_change}. }
\label{fig:event}
\end{figure}

A complicating factor is the unknown long-term stability of the etalon itself. As has been shown in the recent literature, Fabry-Perot etalons can exhibit chromatic drifts, likely due to minute changes in their mirror coatings. E.g., \citenum{Terrien} has demonstrated the chromatic drift of the etalon used for HPF\cite{HPF}, \citenum{ESPRESSO_etalondrift} analyzed the drift of the etalon used on ESPRESSO\cite{pepe2020}. The etalon used for NEID\cite{NEID} shows strong chromatic drifts as well (Sam Halverson, private communication). While we can exclude chromatic drifts of the MAROON-X etalon calibrator exceeding the ones reported for NEID or HPF, we can not exclude chromatic drifts at the level found for ESPRESSO. The limited long-term intrinsic absolute instrument stability of MAROON-X makes a quantification and subsequent removal of this effect challenging. 

Our intermediate solution is to calculate RV offsets specific for different spectral types for each run from the ThAr data, standard stars, and well-characterized science targets. Since the stellar RV information is differently distributed in wavelength depending on spectral type, different targets classes experience different long-term RV drifts based on the chromatic instrumental and calibrator drifts. The uncertainty in these run offsets ranges from 0.5--1.5\,m\,s$^{-1}$. The impact of this uncertainty on science observations spanning many months to years depends on the intrinsic velocity precision of the data and the expected coverage of orbital periods. Measurements aiming at sub-m\,s$^{-1}$ precision for orbits with weeks to months timescale are most affected while orbital periods shorter than the average length of an observing run (2 weeks) are much less so. 

A viable permanent solution is to improve the intrinsic instrument stability of MAROON-X and to augment the calibration scheme with a Laser Frequency Comb (LFC), replacing the ThAr as the absolute frequency reference.

Given the larger than expected temperature dependence of the instrument, a low hanging fruit to improve the RV stability on an absolute level are improvements in its temperature stability. While the active air temperature control of the thermal enclosure housing the spectrograph performs very well, diurnal and seasonal temperature variation find their way to the spectrograph mainly by conductive paths through the chamber floor and the corrugated SS vacuum hose connecting the vacuum tank of the spectrograph inside the thermal enclosure with the turbo pump station outside. 

Stopping the forced venting of the Pier Lab at the end of July of 2020 reduced the seasonal changes and greatly suppressed the diurnal temperature variations\cite{mx2020} from roughly 15\,mK to about 5\,mK daily P-V on the outside of the vacuum tank. Yet weather-driven changes in the passively vented Pier Lab would still lead to temperature excursions of up to 30\,mK over timescales of weeks, which found their way to the optics with little damping. 

In October 2021 we started to establish a temperature controlled active airstream inside the Pier Lab with a temperature stability of 10\,mK rms. This further lowered temperature fluctuations on all timescales. We now have reached a level of 10\,mK P-V over 30 days on the outside of the vacuum tank and less than 5\,mK P-V on the optics inside the tank. Diurnal fluctuations are significantly lower than that. The average instrument drift over a night decreased from the 50\,m\,s$^{-1}$ level seen during commissioning in December 2020 to $\sim$10\,m\,s$^{-1}$ per night during the latest observing runs.

As the next step to further improve the stability of the instrument, we plan to tie the control loop of the thermal enclosure to the average spectrometer temperature instead of the air temperate inside the thermal enclosure. 

Already discussed are measures to lessen the impact of observatory wide power and coolant supply failures as well as more stringent software interlocks and monitoring systems for the MAROON-X hardware to keep the instrumental profile of MAROON-X as stable as possible.

Most importantly, the long-term RV precision of the instrument can only match the excellent short-term precision with a calibrator stable at the 10\,cm\,s$^{-1}$ level over timescales of months to years and the information density to resolve chromatic features in the drift of the etalon. With the current uncertainties involving the long-term panchromatic stability of Fabry-Perot etalons, we decided to upgrade MAROON-X with a Menlo System astrocomb, thanks to a generous grant by the Gordon and Betty Moore Foundation. We anticipate the delivery of the LFC at the observatory by the spring of 2023. Thanks to the fiber-based calibration unit of MAROON-X, additional input ports for calibration sources are readily available. A hybrid approach of simultaneous etalon and nightly LFC exposures should provide an operational scheme to maximize the lifetime of the broadening fiber of the LFC and minimize operational impacts during LFC outages. 

\section{Future Interventions and Upgrades}
In addition to the installation of the LFC, we are currently working on a number of upgrades to further improve the performance and capabilities of MAROON-X: 
\begin{enumerate}
  \item Installation of a solar calibrator to feed disk-integrated sunlight to the spectrograph. Originally planned for 2021, the observatory has finally received approval for the installation of the solar calibrator on the roof of the support building. The enclosure has been delivered to the observatory and work for the  installation are planned for the second half of 2022. We currently anticipate the solar calibrator feed operational at the same time as the LFC.
  \item Chromatic effects in the barycentric correction, in particular extinction driven differences in exposure midpoint between the blue and red arm need to be addressed. The current exposure meter will be augmented with a chromatic version in 2023.
  \item Installation of a low-res spectrometer for the large-FOV sky background fiber. With a number of community science programs concentrating on faint solar type stars, the impact of scattered moonlight on the RVs has become an issue. The standard sky fiber of MAROON-X has about the same efficiency as a single science fiber and does not provide enough SNR on the faint, broadband sky background to adequately measure the contamination level. This problem was anticipated and a large-area fiber (\SI{500}{\micrometer} diameter, $4''$ on sky FOV) was included in the fiber focal plane at the FIU to feed a high-efficiency low-resolution spectrometer. Based on this spectrum, a high-resolution solar template spectrum can be scaled and subtracted from the science spectrum to correct for the solar contamination. We anticipate the construction and installation of this spectrometer in 2023.
\end{enumerate}

\acknowledgments 
 The University of Chicago group acknowledges funding for the MAROON-X project from the David and Lucile Packard Foundation, the Heising-Simons Foundation, the Gordon and Betty Moore Foundation, the Gemini Observatory, the NSF (award number 2108465), and NASA (grant number 80NSSC22K0117).
 
 Based on observations obtained at the international Gemini Observatory, a program of NSF’s NOIRLab, which is managed by the Association of Universities for Research in Astronomy (AURA) under a cooperative agreement with the National Science Foundation on behalf of the Gemini Observatory partnership: the National Science Foundation (United States), National Research Council (Canada), Agencia Nacional de Investigaci\'{o}n y Desarrollo (Chile), Ministerio de Ciencia, Tecnolog\'{i}a e Innovaci\'{o}n (Argentina), Minist\'{e}rio da Ci\^{e}ncia, Tecnologia, Inova\c{c}\~{o}es e Comunica\c{c}\~{o}es (Brazil), and Korea Astronomy and Space Science Institute (Republic of Korea).

\bibliography{report} 
\bibliographystyle{spiebib} 

\end{document}